\documentclass{vospwks2007}

\title{3D Spectroscopy and the Virtual Observatory}
\author{Bryan W. Miller}
\affil{Gemini Observatory}

\usepackage{graphicx}
\begin{document}


\maketitle

\begin{abstract}
Integral field, or 3D, spectroscopy is the technique of obtaining
spectral information over a two-dimensional, hopefully contiguous,
field of view. While there is some form of astronomical 3D
spectroscopy at all wavelengths, there has been a rapid increase in
interest in optical and near-infrared 3D spectroscopy. This has
resulted in the deployment of a large variety of integral-field
spectrographs on most of the large optical/infrared telescopes. The
amount of IFU data available in observatory archives is large and
growing rapidly. The complications of treating IFU data as both
imaging and spectroscopy make it a special challenge for the virtual
observatory. This article describes the various techniques of optical
and near-infrared spectroscopy and some of the general needs and
issues related to the handling of 3D data by the virtual observatory.
\end{abstract}

\section{Introduction}

A basic goal of observational astronomy is to obtain the most
complete, unbiased description of the objects under study. A complete
description of an astronomical observation is inherently
multi-dimensional, at each position on the sky one can measure
intensity, wavelength or energy, and up to four polarization (Stokes)
parameters. Each of these quantities can also vary with time. Since
almost all astronomical objects have properties that are complex ---
structures on all scales, asymmetries, gradients, and variability ---
there is a strong need to obtain full 2D (spatial) information of all
observables. The maximizes the information content of the observations
and provides the most constraints possible for comparison with models
and theories.

However, the nature of light and the technological limitations of
instruments and detectors in general prevent us from making the
``perfect'' observation. Radio and sub-mm interferometric observations
come the closest to the ideal. A map of the UV plane provides
intensity, frequency (energy), and polarization information. The data
cube does contain time information as well, though usually repeated
observations are often needed to study variability. Many modern 2D
X-ray detectors are also inherently multi-dimensional since the time
and energy of each photon are recorded.  However, obtaining high
energy resolution usually means the loss of one spatial dimension.

Likewise, in the optical and near-infrared (NIR) the available
technology limits the number of dimensions that can be observed
simultaneously. Sensitive photon-counting systems such as
photo-multiplier tubes, avalanche photo-diodes, and superconducting
tunnel-junction devices \citep{debruijne02} are ideal for studying
intensity variability on short timescales, but they have low energy
resolution and are either single-channel devices or small arrays and
so are inefficient for studying large areas of sky or for doing
high-resolution spectroscopy. All large-format detectors such as
photographic plates and CCDs are two dimensional and while they are
good at counting the number of photons they have poor intrinsic energy
resolution. Therefore, optics must be used to separate different
wavelengths. Simply placing a dispersive element in the light path is
one valid solution. While this ``slitless'' spectroscopy can be a
useful way of obtaining spectral information over a full 2D field, it
has the disadvantages of high background noise in ground-based
observations, spectral resolution dependent on image quality, and
confusion in crowded regions.  Slitless spectroscopy has been used
extensively to search for emission lines in limited wavelength regions
\citep[e.g.][]{salzer00,douglas02}. The background and spectral
resolution problems are often solved by isolating the region of
interest with a slit; spectra are obtained only along one physical
dimension and the second dimension of the detector contains the
wavelength information. This is very appropriate for compact isolated
objects, but mapping large areas of extended, complicated sources with
a slit is time consuming and it can be difficult to register spatially
dithered spectra. Also, obtaining spatially contiguous spectra is
usually not possible.

However, in the last 30 years clever optical/NIR instrument designers
have developed 3D ($x,y,\lambda$) techniques that produce spectra of
moderate to high resolutions ($R > 100$) over a contiguous
field. Other advantages of these techniques are easier acquisitions
(no need to center precisely in a slit), radial velocities that are
not affected by slit-centering errors, and no light losses that are
otherwise unavoidable with narrow slits. These techniques were
considered rather specialized at first but in the last 10 years the
advantages of the techniques and the successes of the early 3D, or
integral-field, spectrographs have made them popular and very
common. Most current and future large optical/NIR telescopes have or
will have integral-field options in their standard instrument suites.

Therefore, IFU (integral-field unit) data is becoming more and more
common in telescope archives Also, large surveys such as SAURON and
Atlas3D \citep{dezeeuw02,krajnovic07} are being carried out now and
even more ambitious projects are being planned (see below). Thus, it
is important for the Virtual Observatory (VO) to support
integral-field data. This is a challenge because it requires a
combination of imaging and spectroscopic techniques. However, since
IFU data is still relatively new and many common reduction packages do
not include good tools for manipulating IFU data, there is an
opportunity for the VO to set standards and create useful tools that
will draw people to the VO.

The remainder of this proceedings will review the different types of
optical/IR spectrographs and the different formats of data that are
being produced. Then some basic requirements for handling and
visualizing IFU data in the VO will be discussed.

\section{Optical/IR 3D Techniques}

The two main types of optical/IR 3D instruments are spectrometers
and spectrographs. Within each type there are a variety of related
designs. This section will review the basic features of the most
common variations. Some other good recent technical reviews
of different 3D techniques are \cite{bennett00} and
\cite{allsmith06rev}.

\subsection{Spectrometers}

The fundamental principle of a spectrometer is that the wavelength
region of interest is scanned in discreet steps and a narrow-band 2D
image is taken at each wavelength, thus building up a stack of 2D
images, or cube. Interference techniques are used to make the
adjustable narrow-band filters.

One of the most common types of 3D spectrometer is the imaging
Fabry-Perot, sometimes called a tunable filter, in which the
interference between two closely spaced and adjustable
partially-transmissive plates creates a narrow bandpass. There are
also a variety of related techniques \citep{bh00}. The main advantage
of the tunable filter is that it can have a very wide field, up to
several arcminutes in diameter. However, the spectral resolution is
relatively low, between 100 and 1000, and often $R$ and wavelength
coverage must be traded off against each other. Examples of this type
of instrument are the Taurus tunable filter that has been used on the
AAT and WHT and GRIF behind the adaptive optics system at the CFHT
\citep{clenet02}. Also, several new tunable filters are being
commissioned or developed, including the Maryland-Magellan Tunable
filter \citep{dressler06}, FTF2 for the Flamingos~2 NIR MOS
spectrograph for Gemini South \citep{scott06}, and the OSIRIS
instrument on the GTC \citep{castaneda02}.

\begin{figure}
\centering
\includegraphics[width=0.95\linewidth]{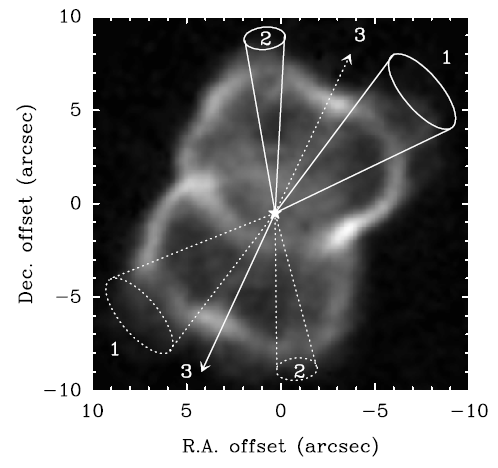}
\caption{A 3D spectrometer result from the imaging FTS BEAR on the CFT
  \citep{cox02}. This shows the velocity-integrated H$_2$ emission
  from the planetary nebula NGC~7027. Outflow axes identified from the
  data cube are labeled.}
\label{fig:n7027}
\end{figure}

One other 3D spectrometer is the Imaging Fourier Transform
Spectrometer (FTS) in which a FTS a Michelson interferometer defines
the bandpass. These instruments have smaller fields-of-view, typically
less than 1 arcminute, but the spectral resolution can be as high as
about 30000. The FTS is sensitive to shot noise from the sky so
ground-based observations will often use a bandpass filter to limit
the spectral range. Figure~\ref{fig:n7027} shows an example image from
the Bear FTS spectrometer on the CFHT \citep{cox02}

The data format of a 3D spectrometer is a regularly gridded 3D
($x,y,\lambda$) datacube. Each ($x,y$) plane is a typical image which
can be manipulated by most image processing software. This data is
similar to a radio datacube and is the easiest to process and
visualize with current software.

The main drawbacks of spectrometers are that the spectral range is
relatively small and, more importantly, that the spectra are not
created simultaneously.  Therefore, any changes in image quality, sky
transparency, and background during the scan will make the spectra
difficult to process and interpret.

\subsection{Spectrographs}

\begin{figure}
\centering
\includegraphics[width=0.9\linewidth]{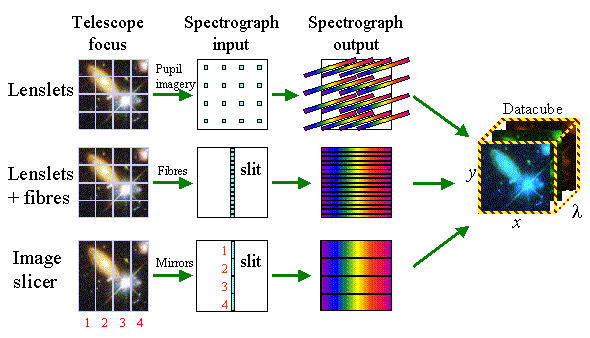}
\caption{A schematic of the three main types of integral field
spectrographs: lenslet-only, lenslet$+$fiber (or fiber only), and
image slicer. The three techniques can produce very similar 3D data but
the information density on the detector increases from top to bottom
(lenslet only to image slicer). (Figure credit: Jeremy
Allington-Smith)}
\label{fig:ifstypes}
\end{figure}

Integral field spectrographs (IFSs) are a more spectra-based approach
to 3D spectroscopy than the spectrometers.  There are three basic
flavors of IFS concepts --- lenslet only, fiber based, and image
slicer --- each with particular advantages and applications
(Figure~\ref{fig:ifstypes}).  In all the flavors optics are used to
sample the focal plane into ``spaxels'' (spatial pixels) and then the
light from each is dispersed by a standard grating or grism
spectrograph. Therefore, the spectral resolution is dependent on the
ruling of the grating/grism and the effective size of the entrance
aperture.  Most IFSs operate with spectral resolutions between 500 and
5000. An advantage of the IFS is that all of the spectra are obtained
simultaneously. However, since detector area is needed to store all
the spectra the field-of-view (FOV) has to be sacrificed. The largest
fields are about an arcminute in diameter. Techniques of modifying the
field size include using fore-optics to change the scale (arcsec/mm),
trading off FOV for spectral coverage, and by increasing the detector
area by using bigger CCDs or CCD mosaics. Some dedicated IFSs exist
but often a multi-purpose spectrograph can be given 3D capability by
inserting an integral-field unit (IFU) into the focal plane as if it
were a standard slit mask.

The first and simplest IFSs were derivatives of the early fiber
multi-object spectrographs (MOS). In this design the input ends of the
optical fibers are packed as closely together as possible in the focal
plan and then the fibers reformat the 2D field into a pseudo slit at
the entrance to the spectrograph \citep{vanderriest80}. Since these
IFUs are relatively easy to construct and incorporate with existing
spectrographs quite a few have seen operation, including SILFID at
CFHT \citep{vanderriest88}, INTEGRAL at the WHT \citep{arribas98}, and
DensPak and SparsPak at WIYN \citep{barden88,bershady04}. The
disadvantages of this design are that the spatial coverage is not
contiguous because of the cladding needed to support the fibers and
that throughput losses occur due to focal ratio degradation (FRD)
caused by feeding the fibers with a slow telescope beam
\citep[see][]{allsmith02}.

\begin{figure}
\centering
\includegraphics[width=0.9\linewidth]{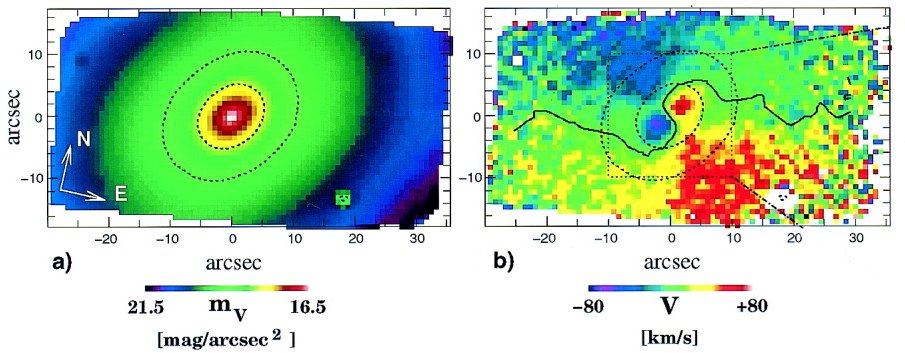}
\caption{The integrated light (a) and velocity field (b) of the inner
  region of the elliptical galaxy NGC~4365 \citep{davies01}. The
  galaxy has a small inner disk rotating around the minor axis but
  stars in the outer galaxy are rotating around the {\rm major} axis.}
\label{fig:n4365}
\end{figure}

The problems of the fiber-only IFU can be avoided by using a
microlense array to sample the focal plane \citep{courtes82}. The
microlenses give full, contiguous spatial coverage and focus an image
of the telescope pupil on the spectrograph entrance. In lenslet-only
IFSs (Figure~\ref{fig:ifstypes}) the spectrograph input is an array of
micropupil images. If grisms are used to disperse the light then the
optical design can be quite simple and have high throughput. However,
a blocking filter must be used to limit the spectral coverage in order to
avoid spectral overlap in the dispersion direction.  Thus, the
spectral coverage is always limited. Also, in the cross-dispersion
direction neighboring spectra overlap at different wavelengths. The
spectra cannot be positioned too closely together and deconvolution is
mandatory to extract the spectra \citep{bacon01}. Field of view is
made adjustable by fore-optics that change the scale (arcsec/mm) at
the lenslet array. Examples of productive lenslet-only IFSs are TIGRE
on the CFHT \citep{bacon95}, OASIS and SAURON on the WHT
\citep{bacon01}, and OSIRIS on Keck
\citep{larkin06}. Figure~\ref{fig:n4365} shows a classic IFS result,
the kinematically decoupled central disk in a velocity map from SAURON
\citep{davies01}.

Lenslet arrays are also used to focus pupil images onto the entrances
of fibers.  Lenslet-fiber IFUs have full spatial coverage and avoid
the FRD losses of fiber-only designs.  Another advantage is that the
spectra can be more tightly packed on the detector since neighboring
spectra on are adjacent on the sky and in wavelength
\citep{allsmith02}. This makes more efficient use of detector area and
helps produce larger fields-of-view. Deconvolution of overlapping
spectra is still helpful, but not mandatory. With IFUs of this type is
often possible to trade field-of-view for spectral coverage. For
example, in the Gemini GMOS IFUs \citep{allsmith02} and the ESO VIMOS
IFU \citep{bonneville03} the full field is divided into equal
sections. Fibers transfer the light from each section to pseudo slits
at the spectrograph(s) entrance(s).  When the largest field-of-view is
required blocking filters must be used to limit spectral coverage and
avoid overlap. When higher spectral resolution or longer wavelength
coverage is desired then some of the pseudo slits can be masked off to
increase spectral coverage at the expense of FOV. This type of IFU has
become very popular.  In addition to the GMOS-N/S and VIMOS IFUs
mentioned above, other examples include CIRPASS on Gemini, the AAT,
and the WHT \citep{parry04}, PMAS on Calar Alto \citep{roth05}, SPIRAL
at the AAT \citep{sharp06}, MPFS on the SAO 6-m, IMACS on Magellan
\citep{schmoll04}, and FLAMES on the VLT \citep{pasquini02}.

The final common type of IFS is the image slicer. In this design the
focal plane of the telescope is sampled by a stack of diamond-turned
slicing mirrors. An optics train then reformats the 2D field ($x,y$)
into a 1D pseudo slit at the spectrograph entrance. The 2D spectra
($x,\lambda$) from the slices are lined up side-by-side on the
detector. Slicer IFU units can be very compact and the all-reflecting
optics make them good for cryogenic environments. Therefore, they are
often used for near and mid-IR IFUs. Slicer IFUs currently in use
include 3D from MPE \citep{weitzel96}, PIFS at Palomar
\citep{murphy00}, UIST at UKIRT \citep{rh06}, ESI at Keck
\citep{sheinis06}, GNIRS and NIFS at Gemini
\citep{allsmith06,hart03,miller06}, and SPIFFI at the VLT
\citep{iserlohe04}.

The native data format for each of these instruments tends to be
unique. For fiber and lenslet IFSs the reduced dataset is a set of 1D
spectra, one for each lenslet/fiber spaxel, and a table of the relative
positions of each spaxel on the sky. Often the lenslets are hexagonal
so this type of data does not naturally produce a rectangularly gridded
3D cube. The slicer IFUs produce a set of ''longslit'' spectra that
are easier to place into a regular cube.

\onecolumn

\twocolumn

\subsection{IFU Surveys}

IFSs are increasingly being used for large surveys. The SAURON
instrument on the WHT was conceived as a survey instrument and the
initial survey of about 70 galaxies has already revealed important new
insight into the dynamics and formation of elliptical galaxies
\citep{emsellem07,cappellari07}. Atlas3D is an extension of the SAURON
survey that will observe nearly 200 early-type elliptical galaxies to
produce a complete sample of early-type galaxies brighter than
$M_K=-21.8$ within 40~Mpc \citep{krajnovic07}.  New IFSs for surveys
are now under development. VIRUS on the HET is a massively-modular IFS
being constructed for the HETDEX galaxy redshift and dark energy
experiment \citep{hill04}. KMOS is a NIR IFS being built for the VLT
which will have 24 slicer units with pickoff mirrors on deployable
arms \citep{sharples06}. It will compile statistics on the kinematics
of distant galaxies similar to what SAURON has done for nearby
galaxies.  Another new IFS for the VLT is MUSE \citep{bacon04}. It
will have 24 spectrograph modules making up a 1 arcminute field. It is
designed to work behind a ground-layer adaptive optics system that
should routinely produce $0.3-0.4$ arcsec image quality. It will be a
general-purpose IFU studying deep fields of distant galaxies, nearby
galaxy kinematics, and solar system objects.

\subsection{IFUs with Adaptive Optics}

If the 1990s was the era when the 8-10 meter became the norm, the
2000s is the time of the common use of adaptive optics (AO) and laser
guide stars (LGS). Such systems are in use or currently under
development at Keck, Gemini, ESO and the GTC they will probably be an
integral part of the next generation of giant telescopes
\citep{crampton06,cunningham06}.  These systems produce
near-diffraction-limited images that rival or surpass the images from
smaller space telescopes. To take advantage of the image quality
spectrographs with AO should have correspondingly narrow
slits. However, normal variations in image quality and the
``breathing'', or plate scale changes normal to single-star AO systems
could lead to significant slit losses with a normal slit spectrograph
IFSs overcome these problems since slit losses are not an issue. IFSs
used with AO systems have small fields, normally around 3 arcseconds,
because of the fine sampling needed, but they provide spectroscopy
while preserving the high image quality. IFSs being using with AO
systems now include OASIS at the WHT, GRIF at CFHT, NIFS at Gemini
North, SINFONI at the VLT, and OSIRIS at Keck. An IFU will also be
used with the high-contrast AO system of the Gemini Planet Imager
\citep{macintosh06}.

\subsection{IFUs on JWST}

IFUs will also be major components of the instrumentation for the
6.5-meter James Webb Space Telescope (JWST) being constructed by NASA,
ESA, and the CSA.  JWST will include three 3D spectroscopy
technologies.  A Fabry-Perot tunable filter to be provided by the CSA
with a 2.2 arcminute field of view will be part of the fine guidance
sensor \citep{rowlands04}. It will provide narrow-band imaging with
$R~100$ and will also have a chonographic mode. Also, the ESA NIRSpec
spectrograph will contain a slicer IFU with a 3 arcsecond
field-of-view and 0.1~arcsec slices \citep{arribas05}. Finally, the
medium resolution spectrometer in the mid-infrared instrument MIRI
is a slicer IFS with four slicers in separate wavelength channels
split by dichroics \citep{wright04}.

\section{3D Spectroscopy and the Virtual Observatory}

The increasing popularity of optical/IR 3D data and the large volume
of 3D data that will soon be in archives makes it very important that
3D spectroscopy is supported by the VO. As with any form of
astronomical data, the main issues that have to be addressed are data
description, storage format, discovery, data transmission, and analysis
tools. 3D data provides an interesting challenge because it is a
mixture of imaging and spectroscopy. Fortunately, this means that
existing protocols and tools for handling imaging and spectroscopic
data separately can serve as a basis for handling 3D data. A goal is
that an astronomer could be able to use the VO to find and combine
multi-wavelength 3D data on an object (a 3D SED) and then do analysis
on this meta-cube. With the addition of multiple observations to give
the time information, we are approaching the goal of the fully general
observation. 

\subsection{Data Description}

A meta-data structure, or general data model, for describing
optical/IR 3D data has been developed using the IVOA Simple Spectra
Access Protocol \citep{chilingarian06}. More details about this
standard and an example application are given in
\citep{chilingarian07}.

\subsection{Storage Format}

The storage format for 3D data is an important consideration for
discovery, transmission, and analysis. As has been mentioned, each IFU
tends to have its own data format. This is partially due to the fact
that the most commonly used data reduction packages such as IRAF,
MIDAS, and IDL, do not have a common data structure for 3D data. Also,
there is the problem that different types of instruments produce
different types of data: stacks of images; one spectrum per spaxel; or
a series of 2D ``longslit'' spectra. A common data format for reduced
data would help promote the use IFU use by making it easier to develop
analysis tools. A common format would also simplify making IFU data VO
compliant.

One option for a standard IFU data format is the Euro3D format
developed by the European OPTICON Euro3D research training network
\citep{euro3d}. The format is a 3D FITS table in which each row
contains the information from one spaxel. Columns can contain
coordinate information and data, data quality, and statistics
vectors. It is general to all IFS instruments after the instrument
signature is removed. It also has a very general spatial description
of a spaxel that does not require resampling to a 3D linear cube and
that supports adaptive binning \citep{cappellari03}.

Another option for a common format is the linearly sampled 3D FITS
cube. This gives the data a similar format to radio and 3D
spectrometer data cubes and it makes the data a bit easier to
visualize with common display tools. However, it would require that
many IFS datasets be re-sampled (which can often degrade data) and it
doesn't lend itself to adaptive binning.  Data that is not on a
regular grid should not be forced into one unnecessarily. However,
because of the meta-data description it should be possible to support
both Euro3D and datacube formats.

\subsection{Discovery}

Data discovery needs to answer the user questions ``What data is
available?'' and ``Is it useful to me?''. If the answer to the second
question is ``yes'' then the astronomer will continue to work on the
data by downloading it or analyzing it with VO tools. For 3D data good
query interfaces must be joined with both imaging and spectral
visualization tools. Searches need to be done both in terms of
position and wavelength. It should be possible to visualize the
footprint of IFU data on a digital sky image. Then it should be
possible to preview both image and spectral representations of the 3D
dataset. A first-generation tool that fulfills many of these
requirements is described by \cite{chilingarian07} in this volume.

\subsection{Data Transmission}

Full Euro3D or datacube files can be many megabytes in size and new
instruments like VIRUS and MUSE will produce vastly larger datasets
than current instruments. Therefore, it is important that it be
possible to extract regions of a dataset. Because of the size and
complexity of 3D data, data providers should experiment with providing
processing and analysis on the server. However, the users should
always have the option to download the data for local analysis.

\subsection{Visualization and Analysis Tools}

\begin{figure}
\centering
\includegraphics[width=0.9\linewidth]{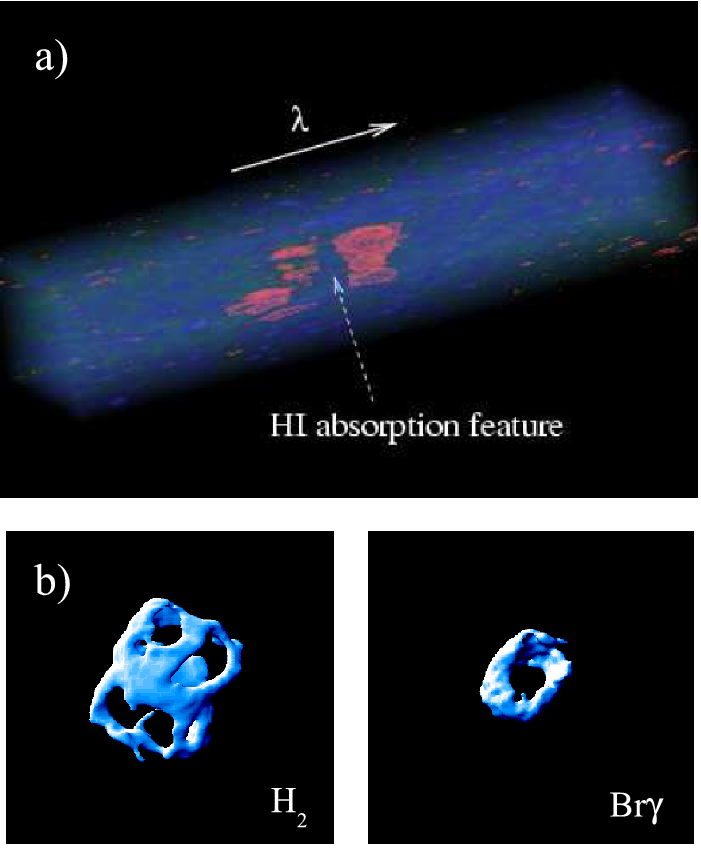}
\caption{Volume renderings of 3D data cubes. a) Ly$\alpha$ emission
  and absorption from a $z=3.09$ galaxy observed by SAURON
  \citep{wilman05}.  b) Representations of the H$_2$ and Br$\gamma$
  emission from the planetary nebula NGC~7027 \citep{cox02}. The
  outflow cavities sketched in Figure~\ref{fig:n7027} are clearly
  seen.}
\label{fig:ifuvol}
\end{figure}

As has been mentioned several times, visualization and analysis of 3D
data is challenging because of the combination of needs for
manipulating images and spectra. It is always important to be able to
treat a 3D dataset as a set of images. It is often useful to loop
through the planes like a movie. For IFS data this means that the tool
needs to reconstruct the images and show any binning that was done.
Registering and either overlaying or overplotting different planes or
different images are also needed. Also, it should be possible to select
a wavelength range and inspect the resulting 2D spatial
map. Inversely, one should be able to select a region from a map and
then inspect either the individual spectra or the summed or averaged
spectrum. As with imaging, overplotting is an important feature. Two
3D visualization tools that do many of these things are the Euro3D
tool \citep{sanchez04} and QFitsView \citep{abuter06}. However,
neither is VO compliant at this time. Another potentially powerful
visualization option is to treat the spatial and spectral dimensions
together. A few visualization labs and authors have been experimenting
with generating 3D views of IFU data by using volume rendering,
stereoscopic views or ``3D'' display monitors, and holographic imaging
(see Figure~\ref{fig:ifuvol}). These techniques make full use of the
data and can help astronomers draw the correct physical
interpretation. Generating good visualization tools for 3D data is a
challenge but also an important opportunity for the VO community.

Until now we have been treating the spectral and spatial domains
independently, however, certain operations on 3D data require that
they be treated simultaneously. One example is the need to subtract a
spectral continuum from a region using an annulus around it. This
would involve fitting in both wavelength and position. Also, both
spatial and spectral PSFs change with wavelength so a deconvolution
of either quantity involves deconvolution in both. For example,
atmospheric refraction is a shift of spatial position with wavelength
and correcting for it requires operations in all dimensions of the
dataset \citep{arribas99}.  Finally, adaptive binning is a means of
combining adjacent spaxels until the spectral signal-to-noise is
sufficient for a particular analysis. All these operations mean that
the errors in the spectra of adjacent spaxels become correlated so
error propagation is very important for the final assessment of
signal-to-noise. 


The analysis needs in only the spectral dimension are fairly standard,
only it must be easy to apply the same analysis to all spaxels in a
dataset. The main needs are single and multiple line fitting, radial
velocity and velocity dispersion measurements using observed templates
or models, and the measurement of line fluxes, equivalent widths, and
line indices. Radial velocity measurements and stellar template
fitting to separate the emission line and absorption line spectra are
good applications for the VO since the template observations or models
could be accessed via the VO. This also meshes with the VO focus on
SED fitting which should be generalized to 3D.


Once a 2D map is produced it should be possible to perform all
standard imaging analysis on it. This includes calculating statistics,
aperture photometry, ellipse or isophote fitting, fitting surfaces,
and the subtraction of background regions. In addition, many maps may
not represent intensity data but the results of the spectral analysis
such as velocity fields or line ratio maps.  These maps may also need
quantitative analysis. An example of this is kinemetry, a Fourier
analysis of line-of-sight velocity distributions similar to ellipse
isophote fitting that provides a quantitative description of velocity
fields \citep{krajnovic06}.

More advanced capabilities can also be envisioned. For example, it is
likely that an investigator would want to apply new and perhaps
proprietary techniques or models to data available with the VO. For
example, someone could want to fit N-body models to velocity fields
from a 3D instrument. This person could download the data to fit to
the models, but if the dataset is very large it may be more efficient
to upload the model or analysis technique to the data on the server
and let the powerful server handle the processing. In cases like this
the intellectual property of the user would have to be respected and
the use of expensive processing resources might need to be negotiated.

\section{Summary}

Three dimensional spectroscopy techniques in the optical and NIR are a
relatively new but rapidly growing in importance. With 3D instruments
on most large telescopes, several large 3D survey instruments
under development, and 3D instruments being prepared for JWST, there
will be en ever growing amount of 3D data in archives which future
investigators will need to access easily in order to take full
advantage of them. Therefore, the VO needs to support the discovery,
transmission, and visualization and analysis of 3D (and higher
dimensional) data. There are challenges to handling these more general
datasets because of the union of spatial, spectral, and eventually
time series and polarization analysis. However, this is also an
important opportunity because tools for handling 3D data are not well
developed. Good tools would draw people with IFU data to the VO and it
would allow the joining of 3D datasets from all wavelength ranges.
The discovery and exploration would allow for the creation of the most
complete combined datasets possible and of data products such as the
spatially resolved SED. The capability to handle such datasets will
encourage data providers to make their data VO accessible and give
investigators more incentive to use the services, thereby maximizing
the potential for scientific discovery.

\section*{Acknowledgments}

The author would like to thank ESAC for the support given to attend
this workshop and Chris Miller for useful discussions about the VO.
The author is also supported by the Gemini Observatory, which is
operated by the Association of Universities for Research in Astronomy,
Inc., on behalf of the international Gemini partnership of Argentina,
Australia, Brazil, Canada, Chile, the United Kingdom, and the United
States of America.

\end{document}